\documentclass{article}
\usepackage{amsfonts}

\input{tcilatex}
\input tcilatex

\begin{document}

\title{Similarity transformations approach for a generalized Fokker-Planck equation}
\author{F. Benamira and L. Guechi \\
Laboratoire de Physique Th\'{e}orique\\
D\'{e}partement de Physique\\
Facult\'{e} des Sciences, Universit\'{e} Mentouri -Constantine\\
Route d'Ain-el-Bey, Constantine 25000, ALGERIA}
\maketitle

\begin{abstract}
By using similarity transformations approach, the exact propagator for a
generalized one-dimensional Fokker-Planck equation, with linear drift force
and space-time dependent diffusion coefficient, is obtained.\ The method is
simple and enables us to recover and generalize special cases studied
through the Lie algebraic approach and the Green function technique.
\end{abstract}

The Fokker-Planck (FP) equation is a very useful tool in modelling a large
number of stochastic phenomena in non equilibrium statistical physics as
well as in chemistry, biology and many other disciplines of science \cite
{Feyn,Risken,Reichl}. The general one dimensional FP equation is given by 
\begin{equation}
\frac{\partial }{\partial t}P\left( x,t\right) =-\frac{\partial }{\partial x}%
\mu _{1}\left( x,t\right) P\left( x,t\right) +\frac{\partial ^{2}}{\partial
x^{2}}\mu _{2}\left( x,t\right) P\left( x,t\right) .  \label{1}
\end{equation}
In this work we are interested in studying the case where $\mu _{1}\left(
x,t\right) $ and $\mu _{2}\left( x,t\right) $ are linear functions of $x$
with time dependent coefficients; namely

\begin{equation}
\mu _1\left( x,t\right) =C(t)x+D(t),\text{ \ and \ }\mu _2\left( x,t\right)
=B(t)+E(t)x.  \label{2}
\end{equation}
Here we assume that the coefficients $B(t),C(t),E(t)$ are arbitrary real
functions of time and $D(t)$ satisfies some constraint that will be stated
later so as to get exact analytic solutions to the problem. Furthermore,
since $\mu _2\left( x,t\right) $ must be positive semi-definite, the
solution has to be limited either to the half axis $x\geq -\frac{B(t)}{E(t)}$
for $E(t)>0$ or to the half axis $x\leq -\frac{B(t)}{E(t)}$ for $E(t)<0.$

By these restrictions (2),\ the FP equation can be written

\begin{equation}
\frac{\partial }{\partial t}P\left( x,t\right) =H(x,t)P\left( x,t\right) ,
\label{3}
\end{equation}
where the operator $H(x,t)$ (hereafter called Hamiltonian) is given by 
\begin{equation}
H(x,t)=-C(t)-\left[ D(t)-2E(t)+C(t)x\right] \frac{\partial }{\partial x}+%
\left[ B(t)+E(t)x\right] \frac{\partial ^{2}}{\partial x^{2}}.  \label{4}
\end{equation}
Note that Eq. (\ref{3}) includes two special cases: the FP equation with a
linear drift force for $E(t)\equiv 0$ which was solved in the framework of
Lie algebraic approach \cite{Lo1} and by the Green function technique \cite
{Demo}, and the generalized FP equation with linear drift force for $%
B(t)\equiv 0$ and $D(t)=\frac{3}{2}E(t)$ \cite{Lo2}$.$

If the coefficients $B,C,D,$ and $E$ are constant, the solution of Eq.(\ref
{3}) may \ be given by the standard method using Laplace of Fourier
transforms of $P(x,t)$ with respect to $x.\;$This leads to a new function,
say $\widehat{P}(k,t),$ that obeys a first order partial differential
equation and can be solved by the method of characteristics$.$

If these coefficients are time dependent function, the solution of Eq. (\ref
{3}) may be written in the form of an ordered integral on time as 
\begin{equation}
P\left( x,t\right) =U(x,t)P\left( x,0\right) ,  \label{5}
\end{equation}
where $U(x,t)$ is the evolution operator. Since, $H(x,t)$ does not commute
with $H(x,t^{\prime })$ for $t\neq t^{\prime },$ $U(x,t)$ reads 
\[
U(x,t)=T\exp \left\{ \int_0^td\tau H(x,\tau )\right\} , 
\]
with $T$ being the time-ordering operator. To get over this difficulty, the
familiar approaches use the Lie algebra \cite{Lo1,Lo2} or the theorem of
Suzuki \cite{Donkov}.

In this paper, we propose to give the solution of this equation \ via the
similarity transformations approach \cite{Fu, Maamache} which consists in
performing a series of transformations on Eq. (\ref{3}) in order to write
the evolution operator as a product of non time-ordering operators whose
their action on $P\left( x,0\right) $ is known.

Performing the transformation $P\left( x,t\right) =X(x,t)$ $Q\left(
x,t\right) ,$ where $X(x,t)$ is an \ arbitrary invertible operator, one gets
the new equivalent equation

\begin{equation}
\frac \partial {\partial t}Q\left( x,t\right) =H^{\prime }(x,t)Q\left(
x,t\right) ,  \label{6}
\end{equation}
with the new Hamiltonian 
\begin{equation}
H^{\prime }(x,t)=X^{-1}(x,t)H(x,t)X(x,t)-X^{-1}(x,t)\frac \partial {\partial
t}X(x,t).  \label{7}
\end{equation}
Notice that since our so-called Hamiltonian is not necessarily Hermitian as
in the quantum mechanical Schroedinger equation, the unitary condition on
the operator $X$ ($X^{\dagger }=X^{-1})$ is not necessary.\ The operator $X$
may have just to be invertible, in this case the transformations are called
similarity transformations.

Our first task is to eliminate the $x\frac{\partial }{\partial x}$ operator
in $H(x,t).$ Thus, we choose $X(x,t)$ as 
\begin{equation}
X(x,t)=\exp \left( -\alpha (t)\right) \exp \left( -\alpha (t)x\frac{\partial 
}{\partial x}\right) ,\text{ \ with }\alpha (t)=\int_{0}^{t}d\tau C(\tau ),
\label{8}
\end{equation}
such that $H(x,0)=H^{\prime }(x,0)$ and $P(x,0)=Q(x,0).$

Since 
\begin{equation}
X^{-1}xX=\text{e}^{\alpha (t)}x,\text{ \ \ \ }X^{-1}\frac{\partial }{%
\partial x}X=\text{e}^{-\alpha (t)}\frac{\partial }{\partial x},  \label{9}
\end{equation}
it is easy to show that 
\begin{equation}
H^{\prime }(x,t)=-\text{e}^{-\alpha (t)}\left[ D(t)-2E(t)\right] \frac{%
\partial }{\partial x}+\text{e}^{-\alpha (t)}\left[ \text{e}^{-\alpha
(t)}B(t)+E(t)x\right] \frac{\partial ^{2}}{\partial x^{2}}.  \label{10}
\end{equation}

When $E(t)\equiv 0,$ and $B(t)>0,$ $H^{\prime }(x,t)$ reduces to a sum of
commuting operators ($\frac{\partial }{\partial x},$ and $\frac{\partial ^{2}%
}{\partial x^{2}}$).\ The equation (\ref{6}) is easily integrated to give 
\begin{equation}
Q(x,t)=\exp \left\{ -d(t)\frac{\partial }{\partial x}\right\} \exp \left\{
b(t)\frac{\partial ^{2}}{\partial x^{2}}\right\} Q(x,0),  \label{11}
\end{equation}
and consequently 
\begin{equation}
P(x,t)=\exp (-\alpha (t))\exp \left( -\alpha (t)x\frac{\partial }{\partial x}%
\right) \exp \left( -d(t)\frac{\partial }{\partial x}\right) \exp \left( b(t)%
\frac{\partial ^{2}}{\partial x^{2}}\right) P(x,0),  \label{12}
\end{equation}
where 
\begin{equation}
\text{\ }b(t)=\int_{0}^{t}d\tau B(\tau )\text{e}^{-2\alpha (\tau )},\text{%
\quad }d(t)=\int_{0}^{t}d\tau D(\tau )\text{e}^{-\alpha (\tau )}.  \label{13}
\end{equation}
Using the well-known relations 
\begin{equation}
\text{e}^{\gamma \frac{\partial }{\partial x}}f(x)=f(x+\gamma ),\text{ \ \ e}%
^{\gamma x\frac{\partial }{\partial x}}f(x)=f(\text{e}^{\gamma }x);\text{ \ }%
\forall \text{ }\gamma .  \label{14}
\end{equation}
as well as the formula 
\begin{equation}
\text{e}^{\gamma \varepsilon \frac{\partial ^{2}}{\partial x^{2}}}f(x)=\frac{%
1}{\sqrt{4\pi \gamma }}\int_{-\infty }^{+\infty }dy\exp \left[ -\frac{1}{%
4\gamma }\left( y-\frac{x}{\sqrt{\varepsilon }}\right) ^{2}\right] f\left( 
\sqrt{\varepsilon }y\right) ,  \label{15}
\end{equation}
valid for $\func{Re}\gamma \geq 0$ and $\varepsilon =\pm 1,$ Eq.(\ref{12})
may be put into the form 
\begin{equation}
P(x,t)=\int_{-\infty }^{+\infty }dyK(x,t;y,0)P(y,0),  \label{16}
\end{equation}
where the transition probability density (propagator) $K(x,t;y,0)$ is given
by 
\begin{equation}
K(x,t;y,0)=\frac{1}{\sqrt{4\pi b(t)}}\exp \left[ -\frac{\left( x\text{e}%
^{-\alpha (t)}-d(t)-y\right) ^{2}}{4b(t)}-\alpha (t)\right] .  \label{17}
\end{equation}
This result agrees with the one obtained in refs \cite{Lo1,Demo}. Notice
that since this is a Gaussian distribution, it may be easily checked by
evaluating the average $\left\langle x\right\rangle $ and the moment $%
\left\langle x^{2}\right\rangle $ directly from Eq. (\ref{1}).

Let us now consider the case where $E(t)\neq 0$. In order to simplify Eq. (%
\ref{6})\ let us transform it by eliminating the $\frac{\partial ^{2}}{%
\partial x^{2}}$ operator in the Hamiltonian $H^{\prime }$ by means of the
unitary transformation defined as 
\begin{equation}
Q(x,t)=\exp \left( a(t)\frac{\partial }{\partial x}\right) R(x,t),
\label{18}
\end{equation}
with. $a(t)=\frac{B(t)}{E(t)}$e$^{-\alpha (t)}.$ It is easy to show that the
Hamiltonian $H^{\prime }$ becomes 
\begin{equation}
H^{\prime \prime }=-\text{e}^{-\alpha (t)}\left( D(t)-2E(t)+\stackrel{\cdot 
}{a}(t)\text{e}^{\alpha (t)}\right) \frac{\partial }{\partial x}+\text{e}%
^{-\alpha (t)}E(t)x\frac{\partial ^{2}}{\partial x^{2}}.  \label{19}
\end{equation}
If we choose $D(t)$ as 
\begin{equation}
D(t)=\frac{3}{2}E(t)-\stackrel{\cdot }{a}(t)\text{e}^{\alpha (t)},
\label{20}
\end{equation}
Eq. (\ref{19}) is reduced to 
\begin{equation}
H^{\prime \prime }=e^{-\alpha (t)}E(t)\left( \frac{1}{2}\frac{\partial }{%
\partial x}+x\frac{\partial ^{2}}{\partial x^{2}}\right) .  \label{21}
\end{equation}
The transformed equation

\begin{equation}
\frac{\partial }{\partial t}R(x,t)=H^{\prime \prime }(x,t)R(x,t),  \label{22}
\end{equation}
is also a Fokker-Planck equation with a diffusion coefficient $e^{-\alpha
(t)}E(t)x,$ which must be positive semi-definite for all $x$ and $t$.\ Thus
the solution of (\ref{22}) must be restricted to the half axis $x\geq 0$ for 
$E(t)>0$ or to the half axis $x\leq 0$ for $E(t)<0.$ This means that for any
time there is no chance for the process to reach the negative half axis for $%
E(t)>0$ or the positive half axis for $E(t)<0.$ Thus $R(x,t)$ is such that 
\begin{equation}
R(x,t)\equiv 0\text{ for }x<0\text{ if }E(t)<0,\text{ or for }x>0\text{ if }%
E(t)>0.  \label{23}
\end{equation}
Setting $E(t)=\varepsilon \left| E(t)\right| $ with $\varepsilon =\pm 1$
being the sign of $E(t),$the formal solution may be put as

\begin{equation}
R(x,t)=\theta (\varepsilon x)\exp \left[ e(t)\left( \frac 12\frac \partial
{\partial x}+x\frac{\partial ^2}{\partial x^2}\right) \right] R(x,0),
\label{22-1}
\end{equation}
where $e(t)=\int_0^td\tau E(\tau )$e$^{-\alpha (\tau )}$.

However, the usual method for solving equations of type (\ref{22}) with the
condition (\ref{23}) is by the use of Green's function technique, via
Laplace of Fourier transforms, and the method of images \cite{Feller}. Here
we shall give an alternative technique that leads to the result using simple
arguments.

Noting that

\begin{equation}
\frac 12\frac \partial {\partial x}+x\frac{\partial ^2}{\partial x^2}\equiv
\frac 14\left( \frac \partial {\partial \sqrt{x}}\right) ^2,  \label{24}
\end{equation}
the solution of the equation (\ref{22}) can be seen as an explicit function
of $\sqrt{x},$i.e. $R(x,t)\equiv $ $\overline{R}(\sqrt{x},t),$ and therefore
if $\overline{R}(\sqrt{x},t)$ is a solution, $\overline{R}(-\sqrt{x},t)$ is
also a solution.\ Thus, the general solution may be written as a linear
combination of the two independent solutions 
\begin{equation}
R(x,t)=a\overline{R}(\sqrt{x},t)+b\overline{R}(-\sqrt{x},t),  \label{25}
\end{equation}
where $a$ and $b$ are determined by the boundary condition at $x=0$ as well
as by the initial condition $\lim_{t\rightarrow 0}R(x,t).$

Reporting (\ref{24}) into (\ref{22-1}) and using (\ref{15}) with (\ref{25})
it is easy to show that $R(x,t)$ corresponding to the conditions (\ref{23})
reads

\begin{eqnarray}
R(x,t) &=&\frac{\theta (\varepsilon x)}{\varepsilon \sqrt{\pi \left|
e(t)\right| }}\int_{0}^{\varepsilon \infty }dz\left\{ a\exp \left[ \frac{-1}{%
\left| e(t)\right| }\left( z-\frac{\sqrt{x}}{\sqrt{\varepsilon }}\right) ^{2}%
\right] \right.  \nonumber \\
&&\left. +b\exp \left[ \frac{-1}{\left| e(t)\right| }\left( z+\frac{\sqrt{x}%
}{\sqrt{\varepsilon }}\right) ^{2}\right] \right\} \overline{R}(\sqrt{%
\varepsilon }z,0)  \label{a26}
\end{eqnarray}
Making the change of variable $y=\varepsilon z^{2},$ (\ref{a26}) reads

\begin{equation}
R(x,t)=\int_{-\infty }^{+\infty }dyG(x,t;y,0)R(y,0),  \label{26-1}
\end{equation}
where $G(x,t;y,0)$ is the transition probability density given by 
\begin{eqnarray}
G(x,t;y,0) &=&\frac{\theta (\varepsilon x)\theta (\varepsilon y)}{2\sqrt{\pi
e(t)y}}\left\{ a\exp \left[ -\frac{1}{e(t)}\left( \sqrt{y}-\sqrt{x}\right)
^{2}\right] \right.  \nonumber \\
&&\left. +b\exp \left[ -\frac{1}{e(t)}\left( \sqrt{y}+\sqrt{x}\right) ^{2}%
\right] \right\} .  \label{26-2}
\end{eqnarray}

This is the general solution of Eq. (\ref{22}) satisfying the conditions (%
\ref{23}). To obtain the physical solution we have to distinguish between
the two boundary conditions at $x=0:R(0,t)=0$ for all $t$ and $R(0,t)\neq 0$
for all $t.$

For the first case, it implies that the transition probability density from
any point $y\neq 0$ to $x=0$ vanishes, i.e. $G(x=0,t;y\neq 0,0)=0$. This
means that the boundary $x=0$ acts as a reflecting barrier. This is
satisfied by choosing $a=-b.$ Furthermore, from Eq. (\ref{26-1}) it is
obvious that $G(x,t;y,0)$ must satisfy the initial condition $%
\lim_{t\rightarrow 0}$ $G(x,t;y,0)=\theta (\varepsilon y)\delta (x-y).$ Then 
$a=-b=1.$

Thus, the physical propagator of Eq. (\ref{22}) with reflecting barrier at
the boundary $x=0,$ that we denote $G^{\text{ref}}(x,t;y,0),$ reads

\begin{eqnarray}
G^{\text{ref}}(x,t;y,0) &=&\frac{\theta (\varepsilon x)\theta (\varepsilon y)%
}{2\sqrt{\pi e(t)y}}\left\{ \exp \left[ -\frac{1}{e(t)}\left( \sqrt{x}-\sqrt{%
y}\right) ^{2}\right] \right.  \nonumber \\
&&\left. -\exp \left[ -\frac{1}{e(t)}\left( \sqrt{x}+\sqrt{y}\right) ^{2}%
\right] \right\}  \nonumber \\
&=&\frac{\theta (\varepsilon x)\theta (\varepsilon y)}{\sqrt{\pi e(t)y}}\exp %
\left[ -\frac{x+y}{e(t)}\right] \sinh \left( \frac{2\sqrt{x}\sqrt{y}}{e(t)}%
\right) .  \label{28}
\end{eqnarray}
For the second case, there is non zero probability for a transition from $%
y\neq 0$ to $x=0.\;$The boundary $x=0$ acts then as an absorbing barrier.
Again from the initial condition $G(x,t;y,0)=\theta (\varepsilon y)\delta
(x-y)$ one gets $a=1.\;$From the formal solution (\ref{22-1}) it is obvious
that if $R(x,0)$ is a non zero constant in the domain of its definition$,$
then $R(x,t)$ does not evolve, i.e. $R(x,t)=R(x,0)$ for all $t>0.\;$To
satisfy this condition, it is obvious from (\ref{26-1}) that $G(x,t;y,0)$
must satisfy the ''backward'' normalization condition

\begin{equation}
\int_{-\infty }^{+\infty }dyG(x,t;y,0)=1.  \label{28-1}
\end{equation}
Integrating (\ref{26-2}) over the variable $y$ for $a=1$ and using (\ref
{28-1}) leads .$a=b=1.$

Thus, the physical propagator of Eq. (\ref{22}) with absorbing barrier at
the boundary $x=0,$ that we denote $G^{\text{abs}}(x,t;y,0),$ reads

\begin{eqnarray}
G^{\text{abs}}(x,t;y,0) &=&\frac{\theta (\varepsilon x)\theta (\varepsilon y)%
}{2\sqrt{\pi e(t)y}}\left\{ \exp \left[ -\frac{1}{e(t)}\left( \sqrt{x}-\sqrt{%
y}\right) ^{2}\right] \right.  \nonumber \\
&&\left. +\exp \left[ -\frac{1}{e(t)}\left( \sqrt{x}+\sqrt{y}\right) ^{2}%
\right] \right\}  \nonumber \\
&=&\frac{\theta (\varepsilon x)\theta (\varepsilon y)}{\sqrt{\pi e(t)y}}\exp %
\left[ -\frac{x+y}{e(t)}\right] \cosh \left( \frac{2\sqrt{x}\sqrt{y}}{e(t)}%
\right) .  \label{29}
\end{eqnarray}
\ 

Now, we turn to the whole solution of our problem. From the relation (\ref
{18}), it is clear that if $a_{0}=a(0)\neq 0$ $(B(0)\neq 0)$, we have 
\begin{equation}
P(x,0)=Q(x,0)=\text{e}^{a_{0}\frac{\partial }{\partial x}}R(x,0),  \label{30}
\end{equation}
where according to (\ref{14}), 
\begin{equation}
R(x,0)=P(x-a_{0},0).  \label{31}
\end{equation}
Since

\begin{equation}
P(x,t)=\exp \left( -\alpha (t)\right) \exp \left( -\alpha (t)x\frac \partial
{\partial x}\right) \exp \left( a(t)\frac \partial {\partial x}\right)
R(x,t),  \label{32}
\end{equation}
we get 
\begin{equation}
P(x,t)=\exp \left( -\alpha (t)\right) R(\overline{x},t),  \label{33}
\end{equation}
with $\overline{x}=(x+\frac{B(t)}{E(t)})$e$^{-\alpha (t)}.$

Substituting Eqs (\ref{28}) and (\ref{29})\ for $x=\overline{x}$ into (\ref
{26-1}) and reporting the result into Eq. (\ref{33}) leads 
\begin{eqnarray}
P_{\text{abs}}^{\text{ref}}(x,t)\! &=&\!\theta (\varepsilon \overline{x})%
\frac{\text{e}^{-\alpha \left( t\right) }}{\varepsilon \sqrt{\pi e\left(
t\right) }}\int_{0}^{\varepsilon \infty }\!\frac{dy}{\sqrt{y}}\text{e}^{-%
\frac{\overline{x}+y}{e\left( t\right) }}\left\{ \!\!\! 
\begin{array}{c}
\sinh \\ 
\cosh
\end{array}
\!\!\!\right. \left( \frac{2\sqrt{\overline{x}}\sqrt{y}}{e\left( t\right) }%
\right) R_{\text{abs}}^{\text{ref}}(y,0)  \nonumber \\
\! &=&\!\theta (\varepsilon \overline{x})\frac{\text{e}^{-\alpha \left(
t\right) }}{\varepsilon \sqrt{\pi e\left( t\right) }}\int_{0}^{\varepsilon
\infty }\!\frac{dy}{\sqrt{y}}\text{e}^{-\frac{\overline{x}+y}{e\left(
t\right) }}\left\{ \!\!\! 
\begin{array}{c}
\sinh \\ 
\cosh
\end{array}
\!\!\!\right. \left( \frac{2\sqrt{\overline{x}}\sqrt{y}}{e\left( t\right) }%
\right) P_{\text{abs}}^{\text{ref}}(y\!-\!a_{0},0)  \nonumber \\
\! &=&\!\theta (\varepsilon \overline{x})\frac{\text{e}^{-\alpha \left(
t\right) }}{\varepsilon \sqrt{\pi e\left( t\right) }}\int_{-a_{0}}^{%
\varepsilon \infty }\!\frac{dy}{\sqrt{\overline{y}_{0}}}\text{e}^{-\frac{%
\overline{x}+\overline{y}_{0}}{e\left( t\right) }}\left\{ \!\!\! 
\begin{array}{c}
\sinh \\ 
\cosh
\end{array}
\!\!\!\right. \left( \frac{2\sqrt{\overline{x}\text{ }}\sqrt{\overline{y}_{0}%
}}{e\left( t\right) }\right) P_{\text{abs}}^{\text{ref}}(y,0),  \nonumber \\
&&  \label{34}
\end{eqnarray}
where $\overline{y}_{0}=y+a_{0}.$

Obviously we may write

\[
P_{\text{abs}}^{\text{ref}}(x,t)=\int_{-\infty }^{+\infty }dyK_{\text{abs}}^{%
\text{ref}}(x,t;y,0)P_{\text{abs}}^{\text{ref}}(y,0), 
\]
with the propagator given by 
\begin{eqnarray}
K_{\text{abs}}^{\text{ref}}(x,t;y,0)\!\!\! &=&\!\!\!\theta \left(
\varepsilon \overline{x}\right) \theta \left( \varepsilon \overline{y}%
_{0}\right) \frac{\text{e}^{-\alpha (t)}}{2\sqrt{\pi \overline{y}_{0}e(t)}}\!%
\left[ \!\text{e}^{-\frac{1}{e(t)}\left( \sqrt{\overline{x}}-\sqrt{\overline{%
y}_{0}}\right) ^{2}}\mp \text{e}^{-\frac{1}{e(t)}\left( \sqrt{\overline{x}}+%
\sqrt{\overline{y}_{0}}\right) ^{2}}\!\right]  \nonumber \\
\!\!\! &=&\!\!\!\theta \left( \varepsilon \overline{x}\right) \theta \left(
\varepsilon \overline{y}_{0}\right) \text{ }\frac{\text{e}^{-\alpha (t)}}{%
\sqrt{\pi \overline{y}_{0}e(t)}}\exp \left( \!-\!\frac{\left( \overline{x}+%
\overline{y}_{0}\right) }{e(t)}\right) \left\{ \!\!\!\! 
\begin{array}{c}
\sinh \\ 
\cosh
\end{array}
\!\!\!\!\right. \left( \frac{2\sqrt{\overline{x}}\sqrt{\overline{y}_{0}}}{%
e(t)}\right) .  \nonumber \\
&&  \label{35}
\end{eqnarray}
$\;$

It is obvious that $\lim_{t\rightarrow 0}K_{\text{abs}}^{\text{ref}%
}(x,t;y,0)=\theta (\varepsilon y)\delta \left( x-y\right) $ as it should be.
However, the ''forward'' normalization condition on the transition
probability density is only satisfied for a process with reflecting boundary
condition:

\begin{equation}
\chi ^{\text{ref}}(\overline{y}_{0},e(t))=\int_{-\infty }^{+\infty }dxK^{%
\text{ref}}(x,t;y,0)=1,  \label{36}
\end{equation}
for all $t>0$ and $y$ in the domain of definition.\ This is obvious since
for reflecting barrier the process never stop and $K^{\text{ref}}(x,t;y,0)$
is a proper probability density. For a process with absorbing boundary
condition, one has 
\begin{eqnarray}
\chi ^{\text{abs}}\left( \frac{\overline{y}_{0}}{e(t)}\right)
&=&\int_{-\infty }^{+\infty }dxK^{\text{abs}}(x,t;y,0)  \nonumber \\
&=&2\frak{R}\left( \sqrt{2\overline{y}_{0}/e(t)}\right) -1+\sqrt{\frac{e(t)}{%
\pi \overline{y}_{0}}}\exp \left( -\frac{\overline{y}_{0}}{e(t)}\right) ,
\label{37}
\end{eqnarray}
for all $t$ and $y$ in the domain of definition, with $\frak{R}(x)$ denotes
the normal distribution. $\chi ^{\text{abs}}\left( \frac{\overline{y}_{0}}{%
e(t)}\right) $ is the probability that the process starting at $y$ does not
reach the barrier before epoch $t$.\ In other words it represents the
distribution of first-passage times at the boundary \cite{Feller}. Obviously
for this case, $K^{\text{abs}}(x,t;y,0)$ is a defective transition
probability density.

The characteristic functions $\varphi _{\text{abs}}^{\text{ref}}$($\rho )$
of $K^{\text{ref}}(x,t;y,0)$ and $K^{\text{abs}}(x,t;y,0)$ with respect to $%
x,$ defined as 
\begin{equation}
\varphi _{\text{abs}}^{\text{ref}}(\rho )=\int_{-\infty }^{+\infty }dx\exp
(i\rho x)K_{\text{abs}}^{\text{ref}}(x,t;y,0),  \label{38}
\end{equation}
may be straightforwardly evaluated. They can be related to $\chi _{\text{abs}%
}^{\text{ref}}$ as 
\begin{equation}
\varphi _{\text{abs}}^{\text{ref}}(\rho )\!\!=\!\!\frac{\exp \left[ i\rho 
\text{e}^{\alpha (t)}\left( \dfrac{\overline{y}_{0}}{1-i\rho e(t)\text{e}%
^{\alpha (t)}}-a(t)\right) \right] }{\left( 1-i\rho e(t)\text{e}^{\alpha
(t)}\right) ^{\frac{3}{2}}}\chi _{\text{abs}}^{\text{ref}}\left( \frac{%
\overline{y}_{0}}{e(t)\left( 1-i\rho e(t)\text{e}^{\alpha (t)}\right) }%
\right) .  \label{39}
\end{equation}
From this expression it is an easy task to obtain the moments of $K_{\text{%
abs}}^{\text{ref}}(x,t;y,0)$ by the usual formula 
\begin{equation}
\left\langle x^{n}\right\rangle (t)=\left. \frac{1}{i^{n}}\frac{\partial ^{n}%
}{\partial \rho ^{n}}\varphi _{\text{abs}}^{\text{ref}}(\rho )\right| _{\rho
=0},  \label{40}
\end{equation}
and check that their corresponding differential equations actually coincide
with those obtained directly from the differential equation (\ref{1}) with (%
\ref{2}) for $K_{\text{abs}}^{\text{ref}}(x,t;y,0).$

Thus we have derived the propagator and its corresponding characteristic
function for a generalized FP equation with time dependent coefficients
satisfying the relation (\ref{20}) such that many special cases can be
easily deduced from it.\ For instance, if we set $B(t)\equiv 0,$ the
propagator obtained in ref.\cite{Lo2} is generalized for arbitrary sign of $%
E(t)$ and the solution is given for the two boundary conditions
corresponding to a reflecting barrier and an absorbing barrier$.$

$\mathbf{Acknowledgements}$

This work is partially supported by a grant for research (2501/24/2000) from
the Algerian government. The authors are very indebted to the referees for
their valuable remarks.

\end{document}